\documentclass[conference]{IEEEtran}
\IEEEoverridecommandlockouts
\usepackage{cite}
\usepackage{amsmath,amssymb,amsfonts}
\usepackage{algorithmic}
\usepackage{graphicx}
\usepackage{textcomp}
\usepackage{xcolor}
\usepackage{lscape}
\usepackage{longtable}
\usepackage{array}
\usepackage{longtable}
\usepackage{colortbl}
\usepackage{url}

\def\BibTeX{{\rm B\kern-.05em{\sc i\kern-.025em b}\kern-.08em
    T\kern-.1667em\lower.7ex\hbox{E}\kern-.125emX}}
\begin{document}

\title{A Secure Open-Source Intelligence Framework For Cyberbullying Investigation\\
%{\footnotesize \textsuperscript{}}
%\thanks{}
}

\author{\IEEEauthorblockN{Sylvia Worlali Azumah}
\IEEEauthorblockA{\textit{School of Information Technology} \\
\textit{University of Cincinnati}\\
Cincinnati, USA \\
azumahsw@mail.uc.edu}
\and
\IEEEauthorblockN{Victor Adewopo}
\IEEEauthorblockA{\textit{School of Information Technology} \\
\textit{University of Cincinnati}\\
Cincinnati, USA \\
adewopva@mail.uc.edu}
\and
\IEEEauthorblockN{Zag ElSayed}
\IEEEauthorblockA{\textit{School of Information Technology} \\
\textit{University of Cincinnati}\\
Cincinnati, USA \\
elsayezs@ucmail.uc.edu}
\and
\IEEEauthorblockN{Nelly Elsayed}
\IEEEauthorblockA{\textit{School of Information Technology} \\
\textit{University of Cincinnati}\\
Cincinnati, USA \\
elsayeny@ucmail.uc.edu}
\and
\IEEEauthorblockN{Murat Ozer}
\IEEEauthorblockA{\textit{School of Information Technology} \\
\textit{University of Cincinnati}\\
Cincinnati, USA \\
ozermm@ucmail.uc.edu}

% \and
% \and
% \and
}

\maketitle

\begin{abstract}
Cyberbullying has become a pervasive issue based on the rise of cell phones and internet usage affecting individuals worldwide. This paper proposes an open-source intelligence pipeline using data from Twitter to track keywords relevant to cyberbullying in social media to build dashboards for law enforcement agents. We discuss the prevalence of cyberbullying on social media, factors that compel individuals to indulge in cyberbullying, and the legal implications of cyberbullying in different countries also highlight the lack of direction, resources, training, and support that law enforcement officers face in investigating cyberbullying cases.
The proposed interventions for cyberbullying involve collective efforts from various stakeholders, including parents, law enforcement, social media platforms, educational institutions, educators, and researchers.
Our research provides a framework for cyberbullying and provides a comprehensive view of the digital landscape for investigators to track and identify cyberbullies, their tactics, and patterns. An OSINT dashboard with real-time monitoring empowers law enforcement to swiftly take action, protect victims, and make significant strides toward creating a safer online environment.  %The proposed interventions can help stakeholders to make significant strides towards achieving this goal.
\end{abstract}

\begin{IEEEkeywords}
Cyberbullying, Open Source Intelligence, Investigation, Data Visualization
\end{IEEEkeywords}

\section{Introduction}
National Center for Education Statistics (NCES) 2019 report indicated that 20.2\% of students aged 12 to 18 experience bullying. Disturbingly, 15\% of these incidents occurred online or via text, highlighting the prevalence of cyberbullying. Moreover, 41\% of students who were bullied at school believed that the bullying would persist. These statistics shed light on the alarming reality of cyberbullying and its potential long-term effects on victims \cite{seldin2019student}.
Research conducted by Armitage et al.~\cite{armitage2021bullying} reveals that bullying has profound and lasting impacts on victims, beginning with mental health issues during childhood. Childhood bullying represents a significant public health concern as it heightens the risk of poor health, social challenges, and educational setbacks during formative years. These consequences extend into adulthood, manifesting as psychopathology, suicidal tendencies, and even involvement in criminal activities \cite{armitage2021bullying}. Cyberbullying, the online counterpart of traditional bullying that has now expanded its reach into the digital landscape, emerged as a pervasive issue with its own set of challenges and consequences. The use of technology is becoming increasingly intertwined with our lives, and addressing the harmful impact of cyberbullying is imperative in ensuring the well-being and safety of individuals in the digital realm.
Cyberbullying has emerged as a growing concern, and it is defined as the intentional and repetitive use of a computer, mobile phone, or other electronic device to cause harm to someone~\cite{hinduja2014bullying}. The term includes sending threats, posting or distributing libelous or harassing messages, and uploading or distributing hateful images or videos that harm others \cite{patchin2006bullies}. Depending on the age of the group and the definition of cyberbullying, a range of 5 to 72 percent of youth experience cyberbullying.
Cyberbullying is a serious problem that affects teens and schools because of its psychological, emotional, behavioral, and physical consequences~\cite{ybarra2007examining}. Cyberbullying has legal implications that law enforcement officers must be aware of and understand. The growth of cell phones and Internet usage among teens has altered conduct and social norms. Cyberbullying is one of the most critical issues facing law enforcement today. According to reports and research, police across the country lack direction, resources, training, and support.
Cyberbullying cases can be complex and time-consuming to investigate, especially on social media, and many law enforcement agencies lack the necessary resources and expertise to handle them effectively \cite{patchin2012preventing}
To help keep law enforcement agents and officers informed, this research seeks to introduce an open-source intelligence pipeline using the Twitter dataset parsed through Splunk to extract and visualize and analyze exploratory data collection to track keywords relevant to cyberbullying research that will be developed in real-time. The importance of this research gives law enforcement officers visibility into cyberbullying data.

\section{Background }
\subsection{Law on Cyberbullying}
The law enforcement agencies need to take a proactive approach to combat cyberbullying. However, current laws controlling the use of electronic devices and social media have fallen behind, making it difficult for law enforcement to handle cyberbullying cases~\cite{law}. In contrast, advanced countries like Canada, India, America, and Australia have already implemented legal measures to address cyberbullying \cite{tagaymuratovna2022cyberbullying}. In the United States, legal differentiation is required between Cyberbullying, Electronic harassment, and Cyberstalking for legal rules to be applied. According to research in \cite{law}, \cite{tagaymuratovna2022cyberbullying}, 19\% percent of American states have bullying laws; 49 out of fifty American states have traditional bullying laws, forty states have laws that address cyber or electronic harassment, and fourteen states have specific laws governing cyberbullying. Therefore 98\% of American states, for example, New Jersey, have established a revenge porn law to prevent the dissemination of nude photos and videos of minors without their consent \cite{matwyshyn2021fake}. Cyberbullying is a criminal offense in Arkansas, North Carolina, Louisiana, and many other states in the United States. Cyberbullies in North Carolina can face a fine of up to \$1,000 or six months in jail.
In Canada, improper behavior on the internet is considered a cyberbullying crime, and provinces have laws or legislation governing cyberbullying. The Canadian Penal Code under section 264(2)(b) defines cyber aggression as frequently contacting one person with another, which can result in up to 10 years imprisonment~\cite{szoka2009cyberbullying}, \cite{tagaymuratovna2022cyberbullying},~\cite{bocij2002online}. 
In Australia, a law passed in New South Wales identifies cyberbullying as a criminal offense, but it only applies when it is committed against an individual or victim \cite{soldatova2012children}. This law also governs all other forms of cyberbullying in the community.
In Ghana, cyberbullying on social media can lead to a minimum of six months and a maximum of three years in prison. However, the UK, Switzerland, and Spain currently do not have specific laws addressing cyberbullying.
The legal procedures for addressing cyberbullying vary across different countries.  %Table I provides a summary of punishments for various forms of cyberbullying in countries that have laws and policies regulated and implemented to curb cyberbullying.

\subsection{Prevalence of Cyberbullying in Social Media}
Determining the prevalence of cyberbullying is challenging as studies have reported vastly different rates. Some studies suggest cyberbullying is a rare occurrence \cite{kowalski2007electronic}, \cite{riebel2009cyberbullying} while others indicate that almost every young person is either a victim or a perpetrator \cite{calvete2010cyberbullying}, \cite{juvonen2008extending}. On average, studies estimate that between 20\% to 40\% of young people experience cyberbullying \cite{tokunaga2010following}.
Patchin et al. \cite{patchin2012cyberbullying} comprehensive literature review found a mean victimization rate of 24.4\% and a mean perpetration rate of 18\% across studies. However, variations in research methodologies, including different definitions of cyberbullying and timeframes studied, likely contribute to the large discrepancy in prevalence rates \cite{tokunaga2010following}. Nonetheless, it appears that cyberbullying is less prevalent than traditional bullying \cite{sourander2010psychosocial}, \cite{wang2009school}. Willard et al.~\cite{willard2007cyberbullying} was among the first to discuss the issue of cyberbullying in literature. They identified various forms of cyberbullying that can occur regardless of the communication medium employed by the bullies in the digital realm. These forms include flaming, harassment, denigration, impersonation, outing, trickery, exclusion, and cyberstalking. Flaming involves heated exchanges using aggressive language, while harassment entails repeatedly sending cruel, insulting, or offensive messages. Denigration involves disseminating derogatory statements about the target electronically, while impersonation involves posing as someone else to damage their reputation or friendships. Outing entails the public sharing of personal information, especially sexual content. Trickery involves deceiving someone into revealing embarrassing secrets, which are then shared online. Exclusion involves deliberately rejecting someone as a friend on social media, thereby creating an "in-group" and "outcast." Finally, cyberstalking involves using electronic communication to harass and threaten another person repeatedly, creating a sense of fear for their safety \cite{popovic2011prevalence}.
P. K. Smith et al.~\cite{patchin2006bullies} conducted a survey of 384 adolescents under the age of 18 who had visited a teen-focused website from different countries. The survey revealed that 11\% of respondents admitted to engaging in cyberbullying, while 29\% reported being victims of online bullying. Similarly, a study conducted in Brisbane, Australia, by Campbell et al.~\cite{campbell2005pilot} surveyed 120 students and found that 11\% of children identified as cyberbullies and 14\% as cyber victims. In Canada, Li's survey of 264 Canadian children in grades 7 to 9 showed that 17\% had cyberbullied others and over a quarter had been cyberbullied \cite{beran2005cyber}. Additionally, a British survey in \cite{popovic2011prevalence} indicated that 25\% of children between the ages of 11 and 19 had experienced electronic bullying or threats. Smith et al.~\cite{smith2006investigation} conducted a study in London where 22\% of the 92 participants between the ages of 11 and 16 reported being victimized by cyberbullying. Kraft et al.~\cite{kraft2006cyberbullying} reviewed 14 studies from Australia, the US, the UK, and Canada and found that cyberbullying victimization rates ranged from 10\% to 42\%, while cyberbullying perpetration rates varied from 6\% to 33\%, with cross-cultural differences.
\begin{figure}[htbp]
\centerline{\includegraphics[width=0.8\linewidth]{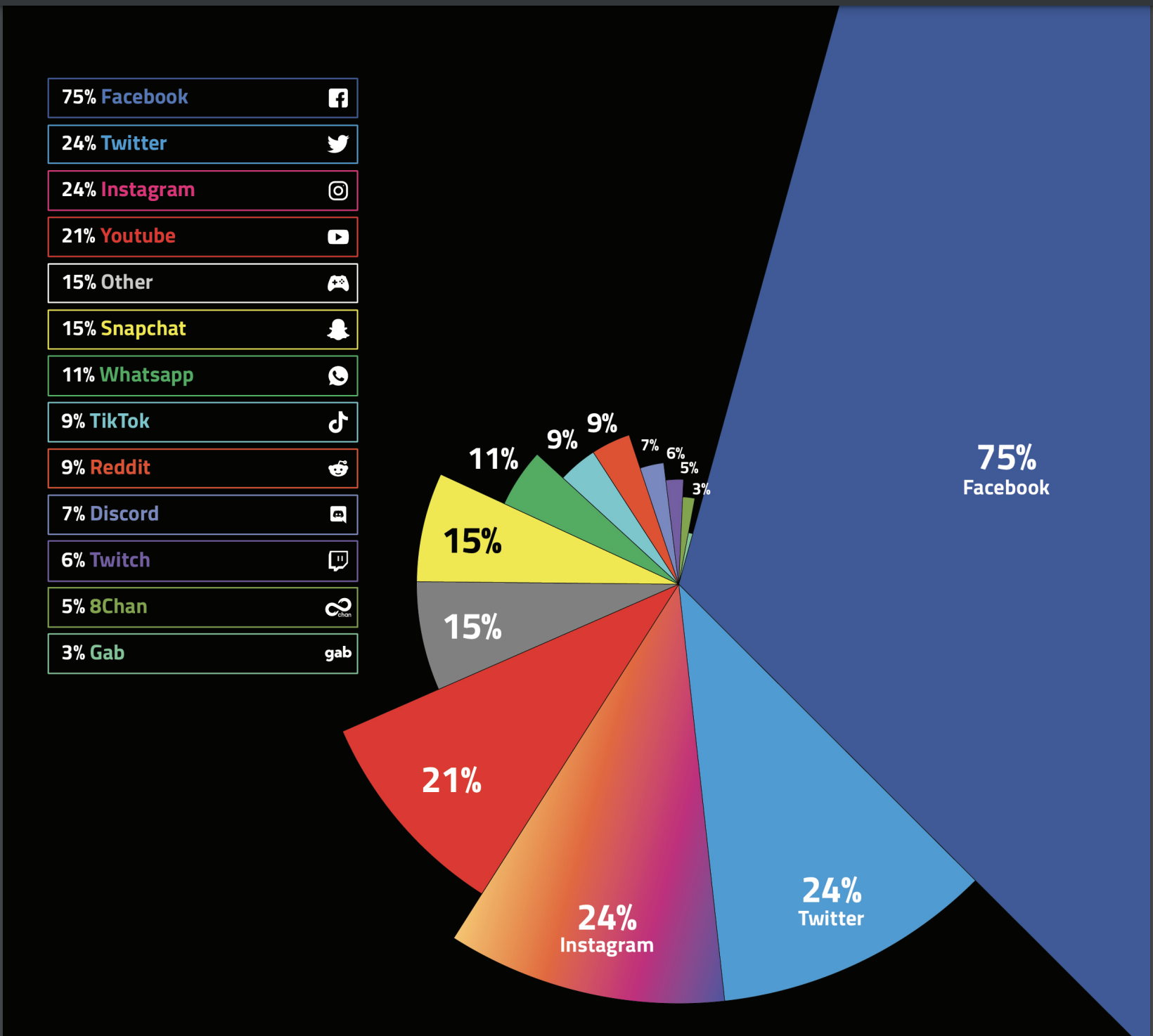}}
\caption{Prevalence of cyberbullying and cyber harassment on online social media platforms \cite{adl}.}
\label{fig1}
\end{figure}
According to recent reports by Anti-Defamation League (ADL) \cite{adl}, Facebook the largest social media platform globally - was found to be involved in the highest proportion of online harassment incidents, with 75\% of victims reporting that at least some of the harassment occurred on the platform. Smaller percentages of individuals reported experiencing online harassment or hate on other platforms such as Twitter 24\%, YouTube 21\%, Instagram 24\%, WhatsApp 11\%, Reddit 9\%, Snapchat 15\%, Discord 7\%, and Twitch 6\%. It is worth noting that Google owns YouTube, while Facebook owns both Instagram and WhatsApp, and Amazon owns Twitch. These findings highlight the need for social media platforms to take proactive measures to address online harassment and ensure user safety. Overall, cyberbullying can take different forms and occur through various media, making it challenging to define and design intervention and prevention programs to combat it.

\subsection{Factors Compelling Individuals to Indulge in Cyberbullying}
There are several factors that can motivate individuals to engage in cyberbullying online, which are worth exploring in more detail. It is important to understand the underlying drivers behind this type of behavior, as they can have serious consequences for both the individual and society at large. By examining these factors, law enforcement can gain a better understanding of the intent behind cyberbullying behavior and how to prevent and address it. The following factors are discussed as follows:
\begin{itemize}
    \item  \textit{Personality:} It is possible that some individuals may possess a dark quad personality, which can lead them to exhibit a lack of empathy towards others and derive pleasure from bullying and intimidating others. This behavior can provide these individuals with a sense of power and control over others, which they may find appealing \cite{chinivar2022online}. It is important to recognize that this type of behavior is not acceptable and can have significant negative impacts on those who are targeted. Strategies should be developed to identify and intervene with individuals who exhibit these personality traits in order to prevent the harm caused by their actions.
    \item  \textit{Mental health:} Frequently, individuals who engage in cyberbullying may be experiencing significant mental health problems as a result of the bullying they have endured. It is important to recognize that these individuals may struggle with serious mental health issues, which can contribute to their online behavior \cite{chinivar2022online}. Addressing the underlying mental health concerns is crucial to address the root causes of this type of behavior effectively. By providing appropriate support and treatment, we can help individuals to overcome these issues and reduce the incidence of online offensive behavior.
    \item  \textit{Boredom:} There is a common saying that "an idle mind is a devil's workshop," which suggests that bored individuals may be more likely to engage in negative or harmful behavior~\cite{chinivar2022online}. This can include cyberbullying behavior, where individuals may experiment with a new persona or engage in conversations that they would not normally consider. It is important to recognize that this behavior can be harmful and can cause significant distress to others. Addressing the underlying issues that contribute to boredom or a lack of purpose can help prevent this behavior and promote healthier online interactions. By providing individuals with positive outlets and opportunities for engagement, we can reduce the incidence of cyberbullying and promote more positive online communities.
    \item  \textit{Loneliness or Exclusion:} There are occasions when individuals may engage in cyberbullying as a result of feelings of loneliness, neglect, or isolation. They may seek to lash out or vent their frustration in this way as a means of coping with their negative emotions \cite{chinivar2022online}. While this behavior is not acceptable, it is important to recognize the underlying drivers of this behavior in order to address it effectively. By providing individuals with support and connection, we can reduce the incidence of this type of behavior and help individuals to develop more positive coping strategies for their negative emotions. It is also important to promote healthy communication and conflict resolution skills to help prevent this type of behavior from arising in the first place.
    \item  \textit{Consequences of breakups or conflict:} IIn some cases, the dissolution of a relationship or friendship due to conflict can give rise to feelings of revenge and jealousy, which may drive individuals to engage in online harassment and offensive behavior. These individuals may seek to harm or intimidate their former partner or friend to retaliate for perceived wrongs or gain a sense of control over the situation~\cite{chinivar2022online}. It is important to recognize that this behavior can be harmful and have serious consequences for both the individual engaging in the behavior and the target of their harassment. By promoting healthy conflict resolution skills and encouraging individuals to seek out positive support networks, we can help to reduce the incidence of this type of behavior and promote healthier online interactions.
    %\item Victims of cyberbullying: Individuals who have experienced bullying may sometimes exhibit offensive behavior as a means of lashing out after being victimized. This behavior can be a response to feelings of powerlessness and helplessness that can result from being subjected to bullying. While this behavior is not acceptable, it is important to recognize that it may be a coping mechanism for individuals who have experienced trauma. By providing individuals who have been victimized with support and resources to help them heal from their experiences, we can reduce the likelihood of them engaging in offensive behavior in the future. It is also important to promote positive communication and conflict resolution skills, as well as to address the root causes of bullying in order to prevent it from occurring in the first place \cite{chinivar2022online}.
\end{itemize}

\begin{figure}[htbp]
\centerline{\includegraphics[width=0.8\linewidth]{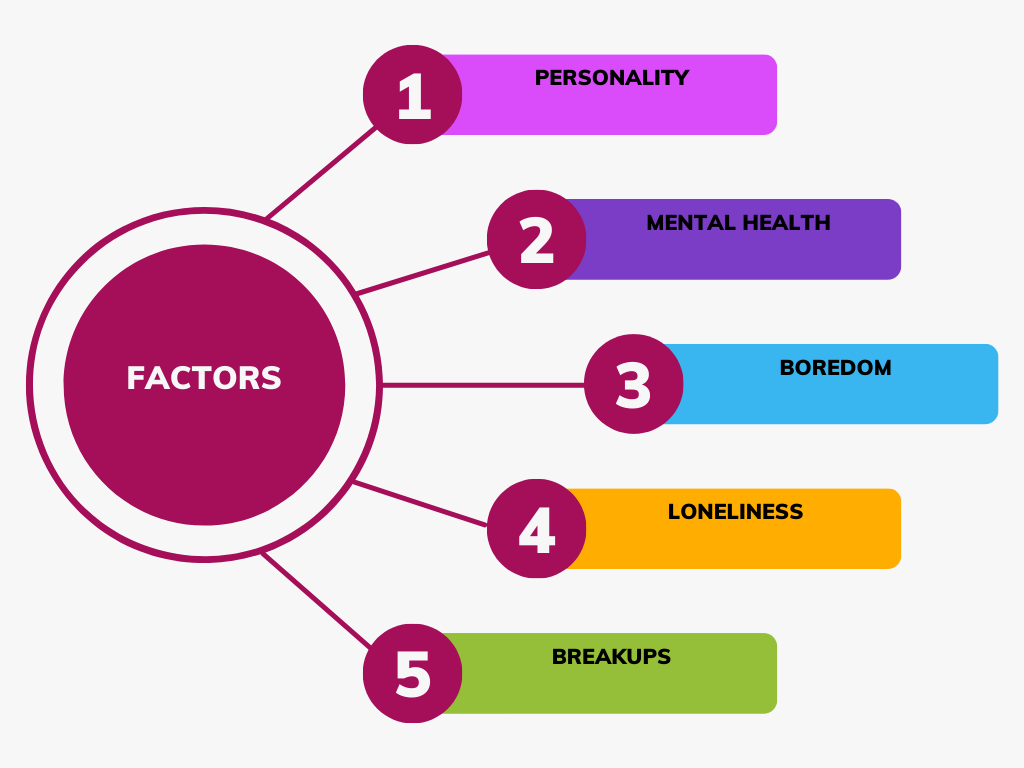}}
\caption{Factors that compel individuals to indulge in cyberbullying on online social networks.}
\label{fig2}
\end{figure}

\section{Method}

This section describes the method of leveraging open-source tools such as Twitter, Twitter API, Splunk, and an existing open-sourced cyberbullying dataset to create a dashboard that aids in the investigation. This involves collecting data from Twitter through the Twitter API, processing the data using Splunk, and creating visualizations and alerts on a dashboard to help with the investigation. The Twitter API can be used to collect data in real-time or historically, depending on the investigation needs. This data can include tweets, user profiles, and location data, among other things.
Figure (\ref{fig}) reveals that Facebook is the primary platform for cyberbullying and harassment, accounting for 75\% of the incidents, followed by Twitter at 24\%. In our research, we focused on utilizing Twitter data due to its accessibility and availability for research purposes through the use of APIs. This allowed us to analyze and gain insights from a significant portion of cyberbullying-related conversations on the platform.

\begin{figure}[htbp]
\centerline{\includegraphics[width=0.8\linewidth]{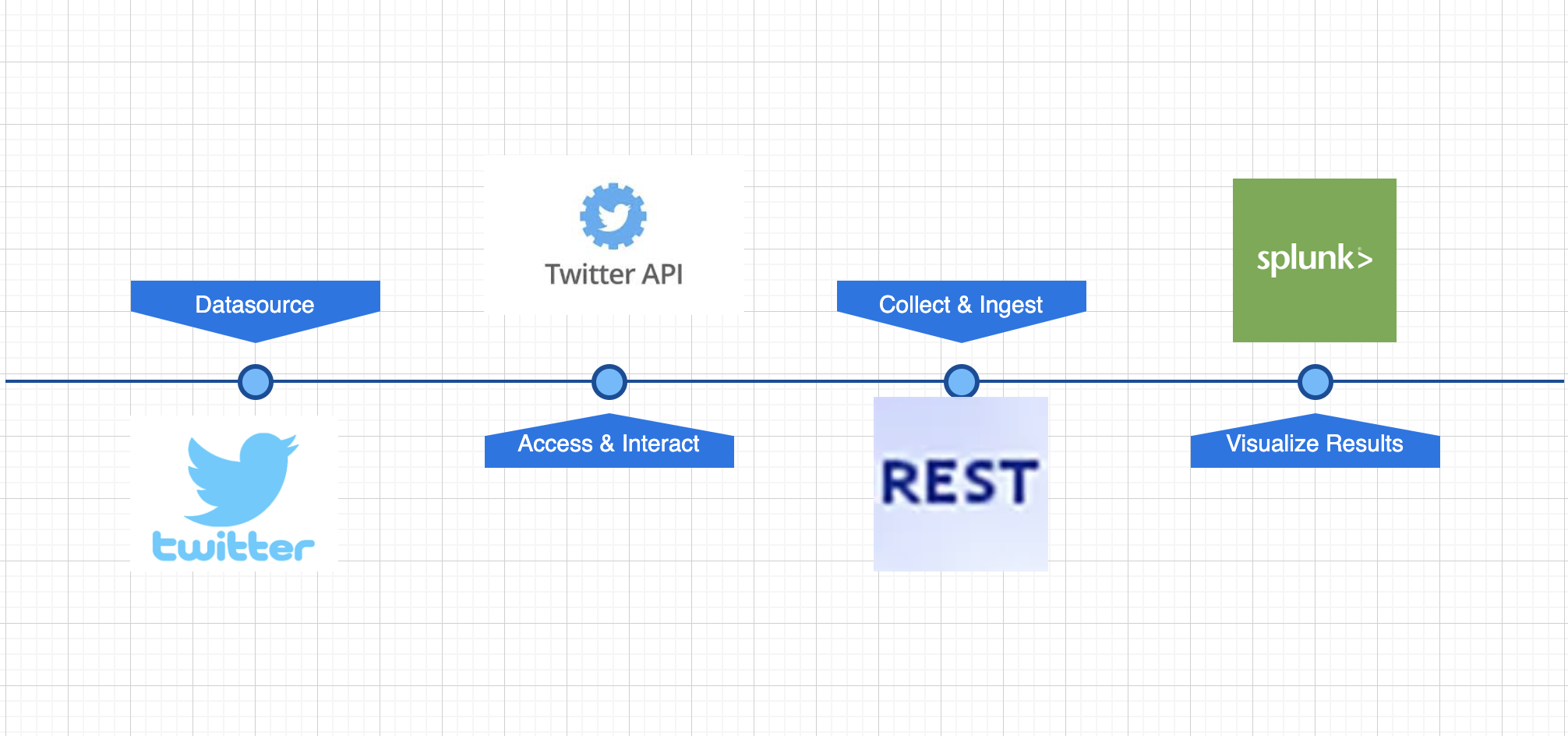}}
\caption{Open source intelligence pipeline for cyberbullying investigation.}
\label{fig3}
\end{figure}
\subsection{Twitter}
Twitter is a social media platform where users can post short messages called "tweets" that are up to 280 characters in length. These tweets can include text, images, and videos and can be seen by anyone who follows the user's account or by anyone who searches for a particular hashtag or keyword.

\subsection{Twitter API}
The Twitter Development API, also called Twitter API, is a collection of programming tools designed to enable developers to interact with the Twitter platform.
With the Twitter API, developers can build applications that can read and write tweets, retrieve user information, and access various other Twitter features, such as trending topics, search results, and notifications.
Multiple versions of the Twitter API are available, each offering distinct features and capabilities. The latest version, v2, grants real-time tweets and other Twitter data access via a RESTful API.
Developers can leverage the Twitter API to develop an array of applications, from social media management tools to data analytics platforms. It is important to keep this in mind. However, that access to the Twitter API is subject to specific terms and conditions. To comply with the Twitter Developer Agreement and Policy, developers must adhere to these guidelines.
\subsection{Rest API}
This Splunk Modular Input is designed to poll data from REST APIs and index the responses. The Python code in this app is compatible with both versions 2.7 and 3.
A REST API modular input is a feature of the Splunk platform that empowers users to gather data from a RESTful web service and incorporate it into Splunk. Modular inputs are a type of input that can be independently configured and managed outside of the main Splunk configuration files. REST API modular inputs allow users to collect data from web services that conform to the principles of Representational State Transfer (REST). To use a REST API modular input in Splunk, a user must provide the RESTful endpoint's URL and any necessary authentication credentials in the input configuration. Splunk then sends a request to the RESTful endpoint, retrieves the data, and indexes it into Splunk.
REST API modular inputs are beneficial for ingesting data from a diverse range of web-based sources, including social media platforms, web applications, and IoT devices. By utilizing RESTful APIs, users can extract structured data from these sources and leverage it for monitoring, analysis, and visualization within the Splunk platform.

\subsection{Splunk}
Splunk is a powerful software platform that empowers organizations to collect, analyze, and visualize data from diverse sources in real time. With its versatile functionalities, Splunk enables users to easily search, monitor, and analyze machine-generated data from servers, applications, network devices, and other systems.
By leveraging Splunk, businesses can gain valuable insights into their operational data, security information, and business metrics, as well as troubleshoot and diagnose problems as they happen. Additionally, the platform can be utilized to track and analyze user behavior, providing organizations with a better understanding of their customers and employees.
Splunk comprises multiple components, including Splunk Enterprise, Splunk Cloud, Splunk Light, and Splunk IT Service Intelligence (ITSI), each equipped with unique features and capabilities. The platform also integrates seamlessly with numerous third-party systems and applications, further expanding its analytical prowess and versatility for businesses.

\subsection{Data Insights}
The existing dataset leveraged in this study is an open-source cyberbullying dataset curated by Ananthi et al.~\cite{Ananthi}. The dataset consists of five types of cyberbullying: sexual harassment, doxing, slut shaming, revenge porn, and cyberstalking. This dataset was selected because it was made publicly available to be replicated by other researchers in the field.
\begin{itemize}
    \item \textit{Sexual Harassment:} According to Harifa et al.~ \cite{harefa2022online}, sexual harassment encompasses behaviors such as sexual comments, unwelcome and inappropriate physical advances, and sexually-oriented conduct in a professional or social setting. It could also be defined as unwanted sexual behavior by an individual or group against others, which is not limited to physical violence or rape. It also includes unwanted sexual advances that can be considered sexual harassment.
     \item \textit{Doxing:} It is a type of cyberbullying that typically takes place on social media platforms like Twitter. This malicious practice involves revealing confidential or sensitive information about someone without their permission. The information that is doxed can include a range of personal details such as demographics, social security numbers, zip codes, GPS coordinates, phone numbers, email addresses, passport numbers, usernames, and passwords \cite{karimi2022automated,waseem2016hateful}. Unfortunately, this type of exposure can lead to a number of negative consequences, such as the formation of hate groups, child trafficking, the spreading of false information or rumors, and even defamation \cite{sari2021persekusi}.
      \item \textit{Slut Shaming:} It refers to a societal process that primarily targets women and involves publicly exposing and shaming them for their perceived sexual availability, behavior, or history \cite{webb2015shame}. Essentially, slut-shaming encompasses the various ways in which women are criticized for their actual, assumed, or imagined sexuality \cite{tanenbaum2015not}. This practice reinforces the idea of sexual virtue, which means conforming to standard sexual behaviors, and upholds the historical tradition of suppressing female sexuality \cite{felmlee2020sexist}. In its most extreme form, slut-shaming involves the dissemination of explicit material with the intention of shaming an individual, as seen in cases of revenge porn, where nude or semi-nude images of someone are shared online by a former romantic partner without the person's consent \cite{webb2015shame}. According to reports, slut-shaming is becoming increasingly common on the internet, specifically online social networks \cite{papp2015exploring}.
       \item \textit{Revenge Pornography:} In many countries, there has been a rise in the distribution of intimate photos or videos without the consent of the person depicted. This phenomenon occurs with and without the consent of the person disseminating the material. It is important to note that the focus is on the distribution of the intimate material without the victim's consent rather than its creation \cite{rosenberg2022revenge}. This non-consensual distribution is frequently driven by revenge after a relationship has ended and often targets women. Websites that promote this behavior typically share photos of women. In court cases where this issue has been addressed, women are generally the victims, and men are the perpetrators \cite{foley2021but}. Due to this gender bias, the phenomenon is commonly referred to as "revenge porn," although the term does not encompass all instances where intimate images are distributed without consent.
        \item \textit{Cyberstalking:} It refers to the use of technology to repeatedly harass, threaten, or communicate with a victim \cite{wilson2022cyberstalking}. Cyberstalkers track and harass their targets across multiple technological platforms. While much of the research on cyberstalking has focused on adolescents and young adults, victims as old as 70 have been identified in some cases \cite{maple2011cyberstalking}.
\end{itemize}

\section{Security}

Ensuring the security of both communication and stored data in the development of an Open-Source Intelligence Dashboard is of paramount importance. One of the primary challenges in securing the dashboard and protecting data from unauthorized access, data breaches, and other security vulnerabilities. To mitigate these risks, we considered some security measures that could be implemented in our system.
\begin{itemize}
    \item \textit{Secure Communication:} Ensuring the security of communication between the Splunk Interface and our data source is the first step established before transmitting data between the Twitter API, Splunk, and the dashboard application. Transport Layer Security (TLS) can be utilized to encrypt data in transit, thus providing confidentiality and integrity during the transmission of data \cite{Rescorla2018TheTL}. We configure the servers and clients in our system to support the latest version of TLS (currently TLS 1.3). This ensures the system uses the most secure and up-to-date encryption methods. We obtain a valid TLS certificate from a trusted Certificate Authority (CA) for the dashboard application's domain to establish secure connections between clients and servers.

    \item \textit{API Integration:} In the connection to the Twitter API and Splunk, we enforce the use of TLS by using HTTPS instead of HTTP for API requests, as HTTPS ensures that the underlying TLS protocol is used for data encryption.

    \item \textit{Access:} The use of API keys and access tokens is essential in authenticating and authorizing Access to the Twitter API. Sensitive credentials are stored in a secure and centralized location using Azure Key Vault, which can prevent unauthorized access and maintain the security of the communication.
    \item  \textit{Data Encryption:} Our data was encrypted at rest and in transit to ensure data integrity using the cipher key tool to encrypt text data. Other at-rest data encryption tools include Transparent Data Encryption (TDE) or Advanced Encryption Standard (AES) algorithms~\cite{Kiraz2016}. These methods ensure that the data stored in the system remains confidential and accessible only to authorized users.
\end{itemize}

To enhance the security of our dashboard application, regular logging and monitoring of user activities and system events are very crucial. Implementing a comprehensive logging system can help identify potential security threats, track user behavior, and support incident response procedures\cite{kent2006sp}. Role-based access control (RBAC) can be employed to grant appropriate permissions to users, ensuring that only authorized personnel can access sensitive data and system functionalities.
Securing our dashboard requires a multi-faceted approach encompassing secure communication channels, data encryption, centralized management of sensitive credentials, robust logging systems, and role-based access control. Implementing these security measures significantly reduces the risk of data breaches and unauthorized access, ensuring the confidentiality and integrity of our dashboard's data collected and analyzed.
\section{Findings}
Industry experts in cybersecurity suggest using social media as a convenient means of staying updated on the most recent security threats, data breaches, and hacks \cite{hoppa2019twitterosint}. Examples of such sources on Twitter include curated content accounts from online forums (such as @Peerlyst), official accounts from security organizations (such as @NISTcyber), and the personal accounts of experts and educators (such as @SchneierBlog and @BrianKrebs) \cite{hoppa2019twitterosint}. It is natural for user-generated online activity to increase during periods when cybersecurity incidents are happening.
The observations made suggested that Twitter data could be valuable for cyberbullying investigations and analysis. 
In this study, Twitter was set up to collect intelligence in this field using a list of terms and concepts that are relevant to the cyberbullying types, including phrases such as slut, sh*t,lo\$er, fat a\$\$, etc. Additionally, a vast corpus of data was included, which was derived from the IEEE Data portal, a research data repository that provides standards-based cyberbullying data. 
The cyberbullying dataset is parsed through Splunk to display all events and content of the dataset.

\begin{figure}[htbp]
\centerline{\includegraphics[width=0.9\linewidth]{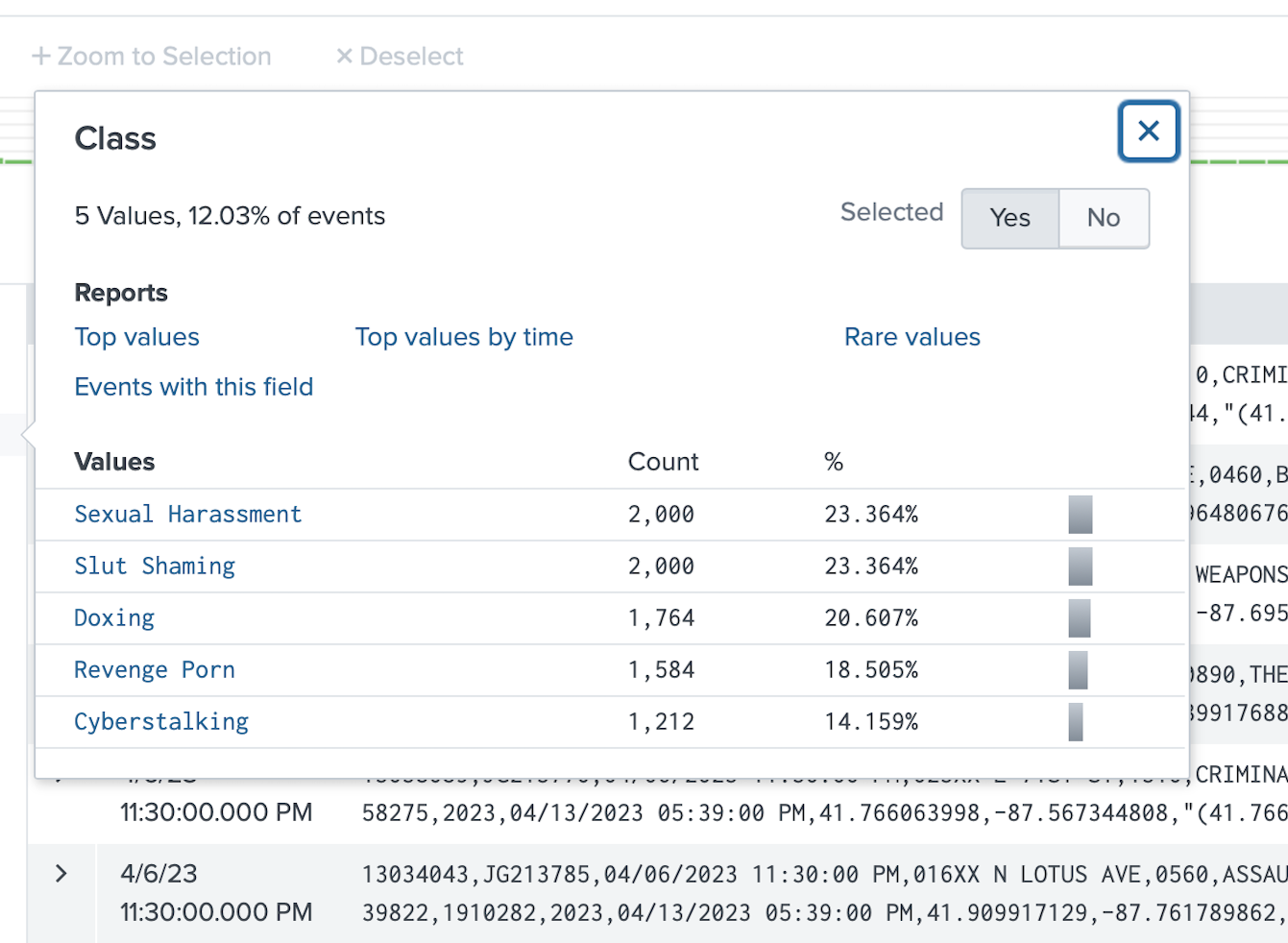}}
\caption{Cyberbullying dataset classes.}
\label{fig4}
\end{figure}

% \begin{figure}[htbp]
% \centerline{\includegraphics[width=0.9\linewidth]{Doxing.png}}
% \caption{Relevant doxing tweets based on keywords.}
% \label{fig}
% \end{figure}

\begin{figure}[htbp]
\centerline{\includegraphics[width=0.9\linewidth]{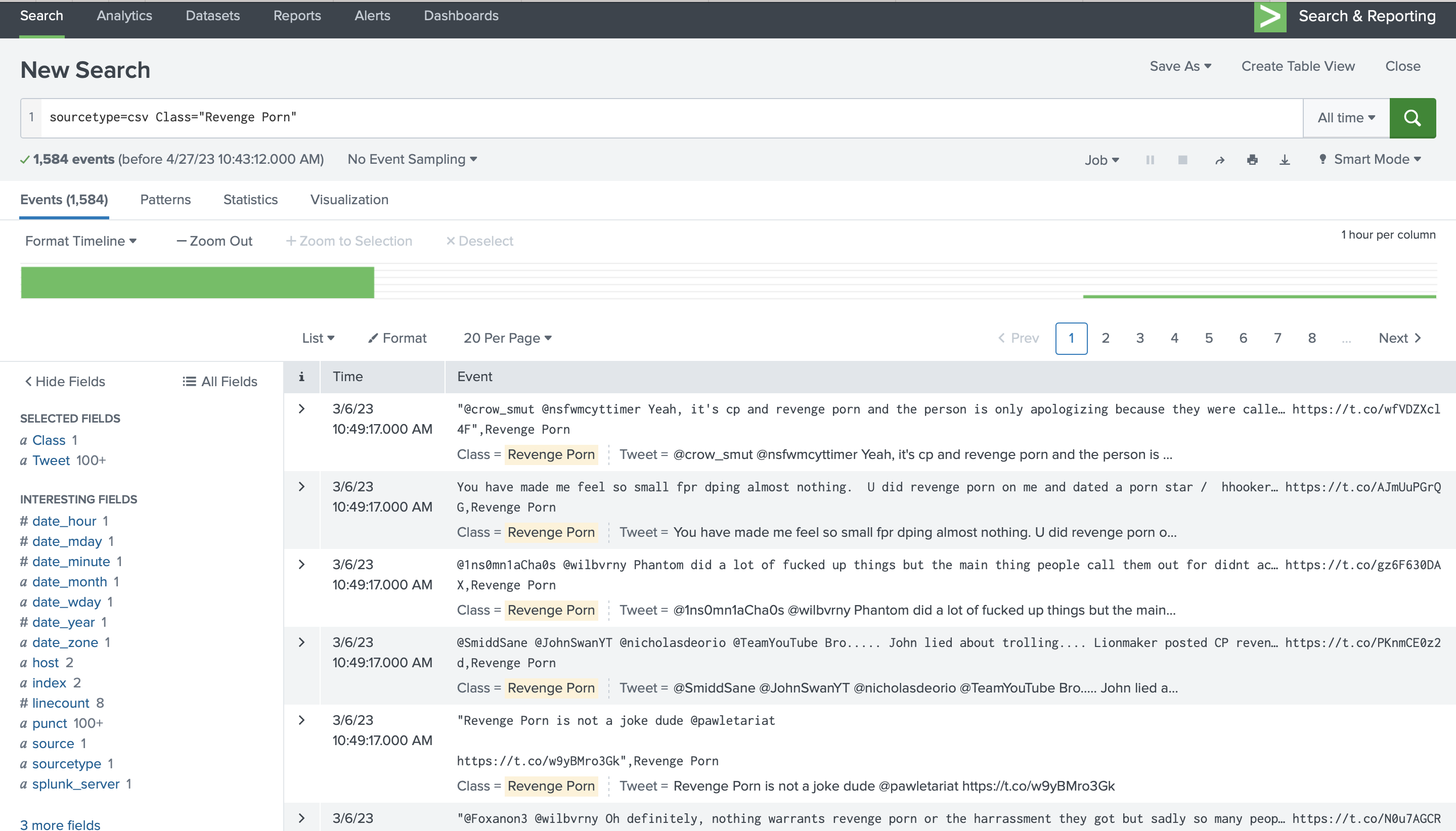}}
\caption{Relevant Revenge Pornography tweets based on keywords.}
\label{fig}
\end{figure}
The excerpt of five types of cyberbullying tweets collected from Twitter using the Twitter API and outputted by Splunk software, based on the configuration described in the method section, is depicted in Figure 4.

% \begin{figure}[htbp]
% \centerline{\includegraphics[width=0.9\linewidth]{SexualH.png}}
% \caption{Relevant Sexual Harassment tweets based on keywords.}
% \label{fig}
% \end{figure}
Figure 5 displays a sample of tweets related to revenge pornography. The data presented in these figures were collected using the Twitter API and indexed using Splunk software. Specific search queries were utilized to extract the relevant tweets, such as "sourcetype=csv Class="Doxing"| rare limit=20" for doxing, "sourcetype=csv Class="Sexual Harassment"| rare limit=20 Tweet" for sexual harassment, and "sourcetype=csv Class="Revenge Porn" for revenge pornography. It should be noted that the tweets included in these figures are limited to those that were selected and indexed during the sourcing process rather than all tweets available in Twitter's full database.

\begin{figure}[htbp]
\centerline{\includegraphics[width=0.9\linewidth]{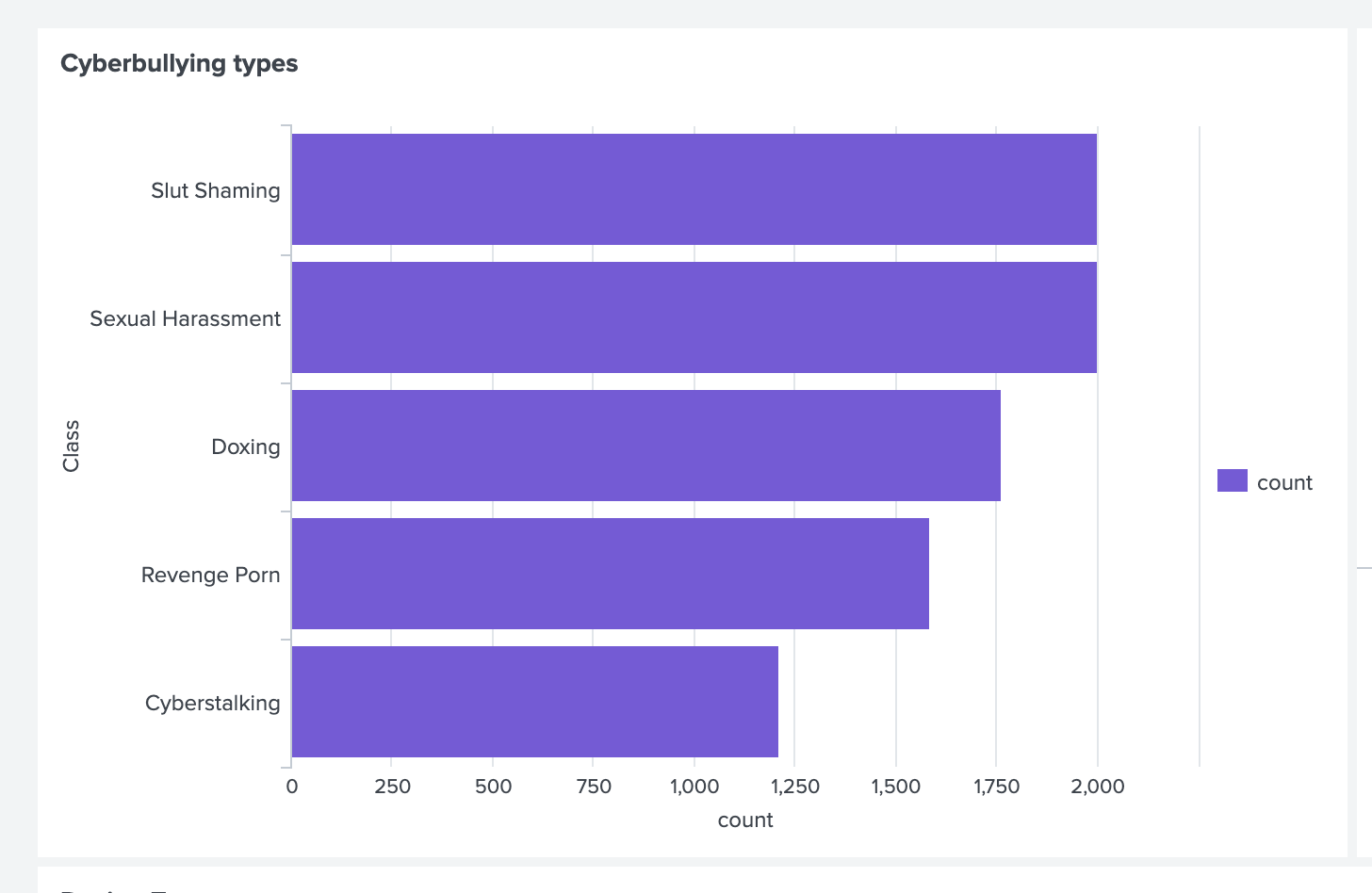}}
\caption{Open source intelligence dashboard for cyberbullying investigation depicting cyberbullying types visualization.}
\label{fig5}
\end{figure}

Despite the limited selection, the amount of data displayed in Figures 4 and 5 may still be overwhelming for a typical human to quickly discern and take action in real time. To address this, alternative visualizations are offered through Splunk Dashboards, which are depicted in Figures 6.
%/Figure 9 and Figure 10 showcase examples of the open-source intelligence dashboard visualizations, which provide a comprehensive overview of the collected data related to cyberbullying. These dashboards present various visualizations that enable easy identification and analysis of the data.

In particular, the count aggregation feature of the dashboard provides an overview of the number of relevant tweets in each cyberbullying class, highlighting the prevalence of each type of cyberbullying. Additionally, the dashboard for doxing and revenge pornography includes a list of relevant tweets, content, or events, along with the specified time period in which they occurred. This enables analysts to quickly identify patterns and trends in the data and take appropriate action to address any issues that may arise.
Overall, these dashboards provide a powerful tool for managing and analyzing large amounts of data related to cyberbullying and can help organizations to develop effective strategies for preventing and addressing this important issue.

% \begin{figure}[htbp]
% \centerline{\includegraphics[width=0.9\linewidth]{Dash2.png}}
% \caption{Open source intelligence dashboard for cyberbullying investigation showing the spikes in the types of cyberbullying searches.}
% \label{fig}
% \end{figure}

\section{Limitation}
Our study is limited by data retrieved from Twitter API calls and also costs associated with the paid version of Splunk Enterprise. Splunk is considerable for larger deployments or organizations with significant data to index. Additionally, the software presents a steep learning curve, potentially necessitating specialized training for effective use. In our research, we used the Splunk Enterprise free version, which is resource-intensive, demanding significant processing power, memory, and applications with a limited amount of data. Furthermore, data retention in Splunk is limited by the license, which may require costly upgrades to extend retention periods.
In the context of the Twitter Development API, there are strict rate limits on API calls to prevent abuse, which can slow down data retrieval or limit the amount of data that can be collected. Another limitation is the potential for changes to the API, which could impact the functionality and usability of existing applications and scripts.

\section{Discussion}
Given the significant societal impact of cyberbullying, it is crucial to focus on methods to prevent and address this issue. Achieving this goal requires a collective effort from various stakeholders, including parents, law enforcement, social media platforms, educational institutions, educators, and researchers, to raise awareness about cyberbullying through knowledge dissemination. Only then can cyberbullying detection research make significant strides worldwide. %/Therefore, the following interventions for cyberbullying are proposed.
Freedom of speech is a fundamental aspect of human existence, serving as a cornerstone that allows individuals to express their thoughts, voice diverse opinions, and contribute to intellectual growth. It plays a vital role in advancing society, fostering progress, and promoting enlightenment. However, it is important to recognize that, like any other right, freedom of speech has its limitations, which are defined by the well-being and dignity of others. In navigating the delicate line between freedom of speech and cyberbullying, the determination of when speech "invades" or "infringes" upon the rights of others is often accompanied by uncertainty due to the subjective nature of interpreting the impact of speech on individual rights \cite{hudson2020cyberbullying}. It is crucial to acknowledge the intricate relationship between these two concepts. While freedom of speech is undeniably essential, its boundaries must be carefully defined to address the harmful effects of cyberbullying. The psychological impact experienced by victims of cyberbullying cannot be overlooked, as it leads to increased levels of anxiety, depression, and a sense of isolation. To address this issue, it becomes imperative to foster a culture that encourages responsible online behavior, promotes digital literacy, and cultivates empathy. Striking a balance between upholding freedom of speech and preventing the adverse consequences of cyberbullying requires a collective effort from society \cite{calvoz2013cyber}. It necessitates the development of an environment that values open dialogue and critical thinking. Education plays a crucial role in equipping individuals with the necessary skills to navigate the digital landscape responsibly. Additionally, platforms and institutions should establish robust policies and mechanisms to address cyberbullying promptly and effectively, ensuring the protection of victims and holding perpetrators accountable. By fostering a culture that promotes responsible online behavior, digital literacy, and empathy, society can effectively navigate the fine line between freedom of speech and cyberbullying \cite{calvert2010fighting}. Such an approach will contribute to the creation of a harmonious digital environment where individuals can freely express themselves while respecting the well-being and dignity of others.
\section{Conclusion}
In conclusion, cyberbullying is a persistent problem that continues to affect individuals, especially teenagers, with its harmful effects on their well-being. This study has presented an open-source intelligence dashboard for cyberbullying investigation, which provides an innovative solution for tracking and analyzing cyberbullying incidents. The dashboard is powered by Twitter data and Splunk software, enabling the extraction, visualization, and analysis of relevant cyberbullying content. Moreover, this study highlights the legal implications of cyberbullying and the lack of resources, training, and support for law enforcement officers to handle cyberbullying cases effectively. The open-source intelligence pipeline presented in this study can help law enforcement officers and other stakeholders to monitor cyberbullying incidents in social media, track trends and patterns, and take appropriate action to prevent and address this issue.
The findings from this study demonstrate the potential of using social media data for cyberbullying investigations and analysis. The open-source intelligence dashboard presented in this study provides a comprehensive overview of the collected data related to cyberbullying, enabling easy identification and analysis of the data. The dashboard offers a powerful tool for managing and analyzing large amounts of data related to cyberbullying and can help organizations to develop effective strategies for preventing and addressing this important issue.
It is crucial to continue exploring new approaches and methods for addressing cyberbullying and improving the safety and well-being of individuals online. With the continued growth of social media and the internet, the issue of cyberbullying is expected to persist. Therefore, it is essential to develop effective strategies and methods for monitoring and preventing cyberbullying incidents. The open-source intelligence dashboard presented in this study provides an innovative solution for addressing this issue, and it is hoped that this research will inspire further work in this area.

\bibliographystyle{IEEEtran}
\bibliography{references}

\end{document}